\documentclass[aps,pra,reprint,showkeys,amssymb,floatfix,longbibliography,superscriptaddress]{revtex4-2}

\usepackage{graphicx}
\usepackage{amsmath,amsfonts,amssymb}
\usepackage{hyperref}
\usepackage{braket}
\usepackage{xcolor}
\usepackage{tikz}
\usepackage{color}
\usepackage{enumerate}
\usepackage[caption=false]{subfig}
\usepackage{multirow}
\usepackage{qcircuit}
\usepackage[normalem]{ulem}
\usepackage{placeins}
\usepackage{url}
\usepackage{dsfont}
\usepackage{braket}
\usepackage[multidot]{grffile}

\newcommand\blue[1]{{\color{black}#1}}

\begin{document}

\title{Garden optimization problems for benchmarking quantum annealers}
\author{\firstname{Carlos D.} \surname{Gonzalez Calaza}}
\thanks{Corresponding author}
\email{c.gonzalez.calaza@fz-juelich.de}
\affiliation{Institute for Advanced Simulation,
  J\"ulich Supercomputing Centre,\\
  Forschungszentrum J\"ulich, D-52425 J\"ulich, Germany}
\author{Dennis Willsch}
\affiliation{Institute for Advanced Simulation,
  J\"ulich Supercomputing Centre,\\
  Forschungszentrum J\"ulich, D-52425 J\"ulich, Germany}
\author{Kristel Michielsen}
\affiliation{Institute for Advanced Simulation,
  J\"ulich Supercomputing Centre,\\
  Forschungszentrum J\"ulich, D-52425 J\"ulich, Germany}
\affiliation{RWTH Aachen University, D-52056 Aachen, Germany}

\date{\today}

\begin{abstract}
  We benchmark the 5000+ qubit system \texttt{Advantage} coupled with the Hybrid Solver Service 2 released by D-Wave Systems Inc.~in September 2020 by using a new class of optimization problems called \emph{garden optimization problems} \blue{known in companion planting}. 
  These problems are scalable to an arbitrarily large number of variables and intuitively find application in real-world scenarios.
  We derive their QUBO formulation and illustrate their relation to the quadratic assignment problem. 
  We demonstrate that the \texttt{Advantage} system
  and the new hybrid solver can solve
  larger problems in less time than their predecessors. However, we also show that the solvers based on the 2000+ qubit system \texttt{DW2000Q} sometimes produce more favourable results if they can solve the problems.
\end{abstract}


\maketitle

\newpage

\section{Introduction}

The quantum processing units (QPUs) of quantum annealers \cite{Johnson2011DWave,Harris2010DWave,Bunyk2014DWave} have doubled in size almost every two years. In September 2020, D-Wave Systems Inc.~has made available the \texttt{Advantage} system \cite{dwave2020Advantage}, having a 5000+ QPU featuring a Pegasus topology with increased connectivity compared to the one of its predecessor, a 2000+ qubit QPU with a Chimera topology in the \texttt{DW2000Q} system \cite{dwave2020TechnicalDescription}. As the complexity of commercially available quantum annealers increases, so does the need for methods to systematically benchmark these systems, using problems which do not become obsolete as quantum annealers grow in size. Therefore, we need problems which are flexibly scalable to an arbitrarily large number of qubits. 

In this paper, we use one such class of scalable problems called \emph{garden optimization problems} to benchmark the \texttt{Advantage} system against the \texttt{DW2000Q} system, as well as the recently released Hybrid Solver Service \verb+hybrid_binary_quadratic_model_version2+ (\texttt{HSSv2}) against its former version \verb+hybrid_binary_quadratic_model_version1+ (\texttt{HSSv1}) and other classical software solvers.

\begin{figure}
    \centering
    \includegraphics[width=\linewidth]{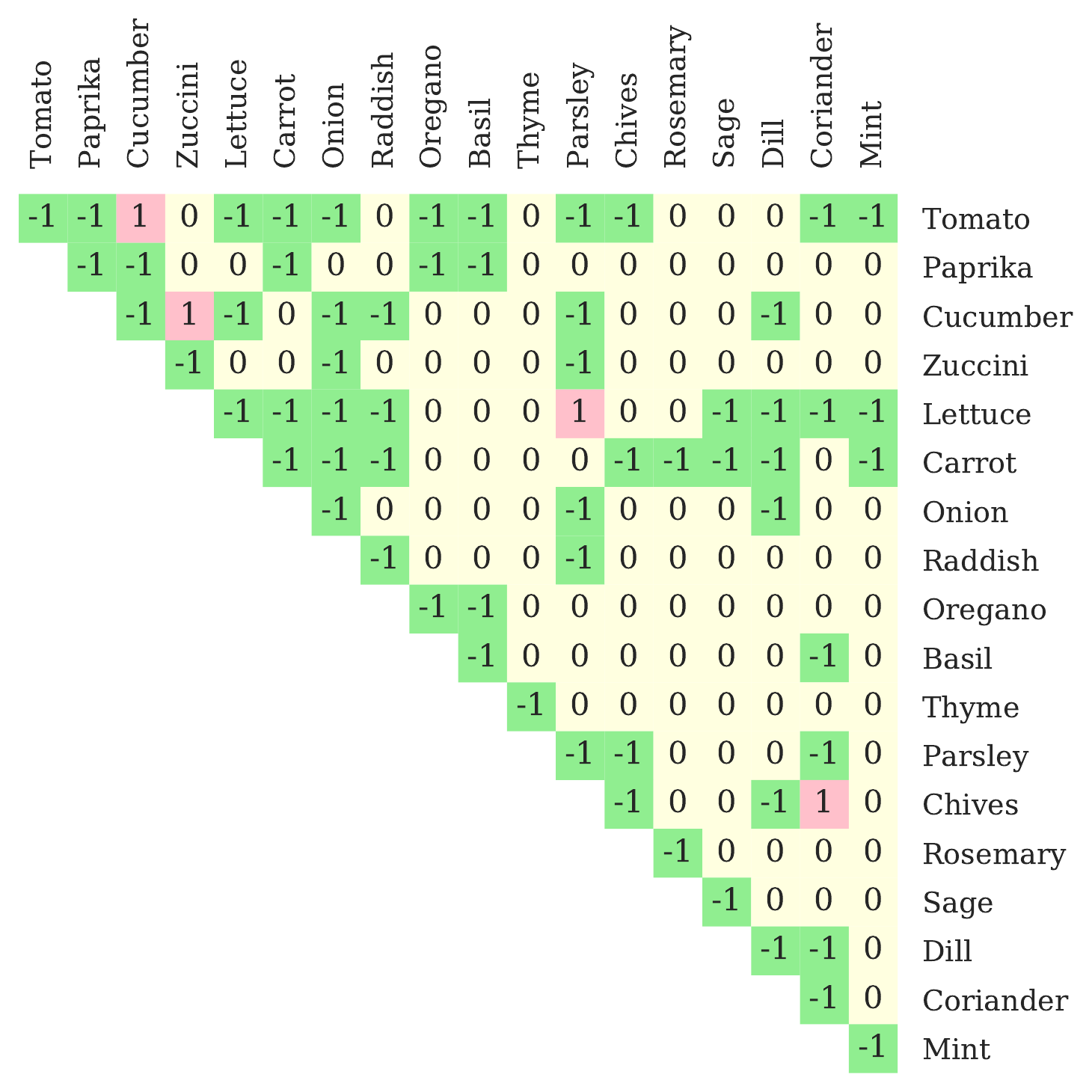}
    \caption{Example of a $t\times t$ companions matrix $C$ for $t=18$ different plant species. The values $C_{jj'}$ of the companions matrix represent friendly (-1, green), neutral (0, yellow), or antagonistic (1, red) relations between different plant species (see Eq.~(\ref{eq:companions})).}
    \label{fig:companions}
\end{figure}

An input problem for a quantum annealer is typically formulated in terms of a quadratic unconstrained binary optimization (QUBO) problem. In this paper, we introduce the QUBO formulation of the garden optimization problem. For this problem, the objective is to find an optimal placement of vegetable plants in a garden, respecting that some plant species have friendly, neutral, or antagonistic relations with other species (see Fig.~\ref{fig:companions})\blue{, a technique known as companion planting}. For instance, tomato and lettuce have a friendly relationship and could be placed next to each other, whereas tomato and cucumber have an antagonistic relationship and should be placed apart from each other.

We argue that the garden optimization problem is well suited to benchmark quantum annealers since it is scalable to an arbitrary number of variables. Furthermore, it represents a problem that finds application in real-world scenarios. Mathematically, the garden optimization problem is closely related to the quadratic assignment problem (QAP) \cite{Koopmans1957QAP,Loiola2007QAPHistoricalSurvey} as well as the constraint satisfaction problem (CSP) \cite{tsang1993foundationsOfCSP}. Such problems have a natural representation in terms of a QUBO problem \cite{Glover2019QuboTutorial}.

We find that the scalability of the garden optimization problem provides the option to benchmark hardware samplers using smaller problem instances as well as hybrid and software solvers by generating larger problem instances. A comparison between D-Wave's \texttt{DW2000Q} and its successor system \texttt{Advantage} using garden optimization problems of up to $100$ variables reveals that, while there were no significant performance differences solving smaller problems, \texttt{Advantage} is capable of embedding and solving much larger problems than \texttt{DW2000Q} (see also the results in Ref.~\onlinecite{Willsch2021BenchmarkAdvantage}). Additionally, we compare the performance of D-Wave's hybrid solvers and some software solvers using problems of up to $12000$ variables. We find that \texttt{HSSv1} returns better results than its successor \texttt{HSSv2} within the same execution time, but is unable to process the biggest problem instances, and \texttt{QBSolv} requires much longer execution times than both hybrid solvers but can return better results than \texttt{HSSv2} for the biggest problems.

The remainder of this paper is structured as follows: In Sec.~\ref{sec:methods}, we introduce the garden optimization problem and formulate it as a QUBO problem suitable as input to quantum annealers. In Sec.~\ref{sec:qa}, we describe the hardware samplers and hybrid and software solvers that are benchmarked in this paper. Section~\ref{sec:results} contains the results of our benchmark study. In Sec.~\ref{sec:conclusion}, we give the conclusions.

\section{The garden optimization problem}
\label{sec:methods}

The goal of the garden optimization problem is to find an optimal placement of $n$ plants into $n$ garden pots (one plant per pot). Each plant belongs to a certain species, and as a matter of fact, some species like to be placed next to each other, whereas others do not. An example of such friendly and unfriendly relationships is shown in Fig.~\ref{fig:companions}. An \emph{optimal} placement of the $n$ plants is thus a placement that maximizes the number of friendly relationships between adjacent pots.

Additionally, we require that each placement respects the following three constraints:
\begin{enumerate}
    \item[(1)] ``Fill all pots'': Each of the $n$ pots shall be filled with exactly one plant.
    \item[(2)] ``Place all plants'': All of the $n$ plants shall be placed in the garden.
    \item[(3)] ``Always look on the bright side of life'': To add an additional degree of complexity, we require that large plant species shall not shadow smaller plants.
\end{enumerate}

In this section, we first describe the mathematical formulation of each of these constraints and the associated cost function that measures an optimal placement, and then rewrite the resulting optimization problem as a QUBO problem.

\subsection{Formulation of the cost function}

Each of the $n$ pots in the vegetable garden is represented by an integer $i=0,\ldots,n-1$. Most generally, the topology of a vegetable garden is defined as an undirected, planar graph $G=(V,E)$, where $V = \{V_i\}$ are the $n$ nodes representing the pots where plants should be placed, and $E = \{E_{i,i'}\}$ are the edges of the graph representing pairs of adjacent pots. In what follows, we only need the adjacency matrix $J$ of $G$, given by
\begin{align}
    J_{ii'} = \begin{cases}
        1 & \text{if $i<i'$ and pot $i$ and $i'$ are adjacent} \\
        0 & \text{otherwise}
    \end{cases}.
\end{align}

A straightforward way to state the problem would be to also enumerate each of the $n$ plants by $n$ integers, and then find an assignment of $n$ plants to $n$ pots. This would lead to a quadratic assignment problem with $n^2$ variables (see Sec.~\ref{sec:QAP} below).
However, as the number of qubits on current quantum annealers is limited, we exploit the fact that for the garden optimization problem, plants can be considered equivalent if they are of the same species. If $t$ represents the total number of species (or plant \emph{types}), this step reduces the number of variables from $n^2$ to $n\times t$.

Therefore, we enumerate the available plant species by an integer $j=0,\ldots,t-1$. The relationship between plant species is encoded in the companions matrix $C_{jj'}$, given by
\begin{align}
    \label{eq:companions}
    C_{jj'} = \begin{cases}
        -1 & \text{friendly relationship} \\
        0 & \text{neutral relationship} \\
        +1 & \text{antagonistic relationship} \\
    \end{cases}.
\end{align}
An example of this matrix for $t=18$ plant types is shown in Fig.~\ref{fig:companions}.

Given the pots $i=0,\ldots,n-1$ and the plant types $j=0,\ldots,t-1$, we define 
$n\times t$ Boolean problem variables $x_{ij}\in\{0,1\}$, with the interpretation that
\begin{align}
    \label{eq:variable}
    x_{ij} = 1 \Leftrightarrow \text{plant of type $j$ is to be placed in pot $i$}.
\end{align}
The cost of placing plants in two connected pots $i$ and $i'$ is thus 
given by $\sum_{jj'} x_{ij} C_{jj'} x_{i'j'}$. Since only one plant shall be placed in each pot (which is to be ensured by constraint (1)), the value of this cost term is ideally equal to $-1$. Shifting this optimal value to zero by adding $+1$
and summing over all adjacent pots for which $J_{ii'}=1$, we arrive at the cost
function of the garden optimization problem
\begin{align}
    \label{eq:objectivefunction}
    \texttt{cost}(\{x_{ij}\}) =
        \sum\limits_{i,i'=0}^{n-1}
        J_{ii'} \left(1+
        \sum\limits_{j,j'=0}^{t-1} x_{ij} C_{jj'} x_{i'j'}\right).
\end{align}

Note that the cost function has been constructed in such a way that it has a lower bound zero. The special value $\texttt{cost}(\{x_{ij}\})=0$ implies that the placement $\{x_{ij}\}$ is optimal, in the sense that all neighbouring plants have a friendly relationship. In this case, we can tell solely from the solution energy that an optimal solution exists and has been found. This is a desirable property for an optimization problem since for a general optimization problem, it is typically not possible to tell if a solution represents the global optimum or not.

However, if no completely friendly arrangement of plants exists for a given problem instance, then the minimum value of the cost function will be larger than zero. In this case, it will not be possible to verify by means of the solution energy that the ground state has been found. Still, the value of the solution energy will give an indication as to how many neutral (weight $+1$) or antagonistic (weight $+2$) neighbourships exist in the produced arrangement of plants.

We remark that this property will continue to hold also after the constraints are included. The reason for this is that, by construction, all constraints will have a positive contribution to the solution energy if they are violated, and no contribution if and only if they are satisfied (cf.~Eq.~(\ref{eq:my_qubo}) below). Thus, after checking the constraints for a given placement $\{x_{ij}\}$, we can interpret the value of the solution energy in the same way as the value of the cost function Eq.~(\ref{eq:objectivefunction}) before. The mathematical formulation of the constraints is the topic of the following section.

\subsection{Formulation of the constraints}

Given the problem variables $x_{ij}\in\{0,1\}$ with the interpretation formulated in Eq.~(\ref{eq:variable}), we can mathematically state the constraints (1)--(3) as follows:
\begin{enumerate}
    \item[(1)] ``Fill all pots'': 
    \begin{align}
    	\label{eq:const_pots}
        \forall i:\quad \sum\limits_{j=0}^{t-1} x_{ij} = 1.
    \end{align}
    \item[(2)] ``Place all plants'':
    \begin{align}
    	\label{eq:const_plants}
        \forall j: \quad \sum\limits_{i=0}^{n-1} x_{ij} = c_j,
    \end{align}
    where $c_j$ denotes the total number of plants of type $j$. Note that by
    definition, we have $\sum_j c_j = n$.
    \item[(3)] ``Always look on the bright side of life'':
    For this constraint, we assign to each plant type $j$ a size $s_j\in\{0,1\}$, 
    where $s_j=0$ ($s_j=1$) means that plant type $j$ is a large (small) species.
    Furthermore, for the sake of concreteness, we fix the topology of the garden to be rectangular such that $i\%2=0$ ($i\%2=1$) represents an even (odd) row in the garden (we use the symbol $\%$ to denote the integer modulo operation). The constraint shall then be fulfilled by placing large plants in even rows and small plants in odd rows, i.e.,
    \begin{align}
    	\label{eq:const_size}
    	\forall i,j: \quad (i\%2 - s_j)^2 x_{ij} = 0.
    \end{align}
    This means that if plant type $j$ is placed in pot $i$ (i.e., $x_{ij}=1$), then 
    we require $i\%2 = s_j$.
\end{enumerate}

\subsection{Relation to the quadratic assignment problem}
\label{sec:QAP}

The garden optimization problem characterized by Eqs.~(\ref{eq:objectivefunction})--(\ref{eq:const_size}) is closely related to the well-known quadratic assignment problem (QAP), which is originally due to Koopmans and Beckmann \cite{Koopmans1957QAP} (see \cite{Loiola2007QAPHistoricalSurvey} for a brief historical survey). The QAP can be formulated as
\begin{align}
    \label{eq:QAP}
    &\text{minimize} & &
        \sum\limits_{ijkp} f_{ij} d_{kp} x_{ik} x_{jp},
    &\\
    &\text{subject to} & &
    \label{eq:QAPconstraint1}
        \forall i: \sum_j x_{ij} = 1,
    \\&&&
    \label{eq:QAPconstraint2}
        \forall j: \sum_i x_{ij} = 1,
\end{align}
where all indices $i,j,k,p$ range from $0$ to $n-1$, the problem variables are $x_{ij}\in\{0,1\}$, and $(f_{ij})$ and $(d_{kp})$ are matrices characterizing the problem instance. The QAP is a difficult combinatorial optimization problem that has been shown to be NP-hard \cite{Sahni1976QAPisNPhard} and can typically not be solved in reasonable time for general instances with $n>30$ \cite{Loiola2007QAPHistoricalSurvey}.

Comparing the garden optimization problem given by Eqs.~(\ref{eq:objectivefunction})--(\ref{eq:const_size}) and the QAP given by Eqs.~(\ref{eq:QAP})--(\ref{eq:QAPconstraint2}), we see that the main differences are: (a) one dimension of the problem variables has been reduced from $n$ to $t$ (cf.~Eq.~(\ref{eq:variable})) to reduce the number of qubits required, resulting in the modified constraint Eq.~(\ref{eq:const_plants}); (b) the minimum of the cost function Eq.~(\ref{eq:objectivefunction}) has been shifted to zero; (c) the difficulty of the problem has been slightly increased by the additional constraint Eq.~(\ref{eq:const_size}), which can be straightforwardly included in the problem's QUBO formulation (see below).

For the QAP, there is a canonical way of obtaining its QUBO formulation (see e.g.~\cite{Glover2019QuboTutorial}). In the following sections, we pursue and extend this approach to obtain the QUBO formulation of the garden optimization problem.

\subsection{QUBO formulation of the problem}
\label{qubo_formulation}

A QUBO problem is defined as the minimization of 
\begin{align}
	\label{eq:general_qubo}
	E(\{y_k\}) = \sum_{k \leq k'}{y_{k}Q_{kk'}y_{k'}}    
\end{align}
where $y_k \in \{0,1\}$ are the binary problem variables, $Q_{kk'}$ is referred to as the QUBO matrix, a real-valued upper-triangular matrix encoding the problem constraints and the objective function, and $E(\{y_k\})$ is the energy to be minimized by finding an appropriate assignment of the variables $y_k$. 

In order to map the problem variables $x_{ij}$ onto the QUBO variables $y_k$, we use the following unary encoding scheme: We have one variable $y_k\in\{0,1\}$ for each node and species combination, so that $n\times t$ gives the total number of variables required for this problem. The index $k$ has to uniquely identify the node $V_i$ and the species $j$, so we compute it from $(i,j)$ via $k = ti + j$ and convert it back to $(i,j)$ using integer division $i=k//t$ and the integer modulo operation $j = k\%t$. The value $y_k=1$ shall represent the statement ``on node $V_i$ there is a plant of species $j$'' (cf.~Eq.~(\ref{eq:variable})). Conversely, $y_k=0$ means ``on node $V_i$ there is no plant of species $j$''. Of course, in a good solution we would expect $n$ of the $y_k$ to be one and $tn-n$ to be zero. 

The garden optimization problem given by Eqs.~(\ref{eq:objectivefunction})--(\ref{eq:const_size}) now needs to be expressed as a QUBO problem, i.e., a minimization of Eq.~(\ref{eq:general_qubo}). To do this, we need to remove the constraints given by Eqs.~(\ref{eq:const_pots})--(\ref{eq:const_size}) by combining them with the cost function Eq.~(\ref{eq:objectivefunction}) into a single expression. This is achieved by adding the constraints as squared penalty terms (whose minima correspond to fulfilling the constraints) to the cost function. The penalty terms are multiplied by Lagrangian multipliers $\lambda_1, \lambda_2, \lambda_3$ which control the importance of each constraint relative to that of the cost function. The values of these multipliers should be chosen to be large enough positive scalars so that the candidate solutions respect the constraints, but not too big that these constraints dominate over the objective function in their contribution to the final energy \cite{Glover2019QuboTutorial}. This way we ensure that valid solutions are returned, while still being able to distinguish between good and poor solutions. We experimented with different values and found that setting the multipliers to $\lambda_1 = \lambda_2 = 2$ and $\lambda_3 = 1$ achieves this goal.

Combining all this, the garden problem stated as a QUBO problem takes the following form:  
\begin{equation}
    \begin{split}
    	\label{eq:my_qubo}
        \min_{x_{ij}\in\{0,1\}}\Bigg\{
        & \sum\limits_{i,i'=0}^{n-1}
        J_{ii'} \left(1+
        \sum\limits_{j,j'=0}^{t-1} x_{ij} C_{jj'} x_{i'j'}\right)\\
        +&\lambda_1
        \sum\limits_{i=0}^{n-1} \left(1 - \sum\limits_{j=0}^{t-1} x_{ij}\right)^2\\
        +& \lambda_2
        \sum\limits_{j=0}^{t-1} \left(c_j - \sum\limits_{i=0}^{n-1} x_{ij}\right)^2\\
        +& \lambda_3
        \sum\limits_{i=0}^{n-1}\sum\limits_{j=0}^{t-1} \left(i\%2 - s_j\right)^2 x_{ij}
        \Bigg\}
    \end{split}
    \end{equation}
By multiplying out the squares, replacing all linear terms $\propto x_{ij}$ by quadratic terms $\propto x_{ij}x_{ij}$ (since $x_{ij} = x_{ij}^2$ for Boolean variables), replacing $x_{ij}$ by $y_k$ and dropping all constant terms, we obtain the final values of the QUBO matrix $Q_{kk'}$ in Eq.~(\ref{eq:general_qubo}). Although the value of the dropped constant terms does not affect the minimum, it is useful to add it to the final energies $E(\{y_k\})$ to maintain the property that the global minimum of the optimization problem has value zero. The explicit construction of the QUBO matrix for the garden optimization problem can be found in Appendix \ref{app:expanded_qubo} and in the example Jupyter Notebook available at Ref.~\onlinecite{GardenOnline}.

\section{Quantum annealing}
\label{sec:qa}

There are two main paradigms in quantum computing: gate-based quantum computing \cite{Benioff1980QuantumTuringMachine,Feynman1982Simulating,Deutsch85QuantumComputer,NielsenChuang} (QC) and quantum annealing \cite{Finnila1994QuantumAnnealing,KadowakiNishimori1998QuantumAnnealing,Fahri2000AdiabaticQuantumComputation,Harris2010DWave,Johnson2011DWave} (QA). They are substantially different in their mode of operation as well as in the current system sizes and the range of problems that they can solve. In theory, gate-based QC can tackle a much wider range of problems due to its universal nature, but current implementations \cite{ibmquantumexperience2016,Google2019QuantumSupremacy} are too small to outperform classical computers in real life applications. On the other hand, QA excels at solving optimization problems such as the QUBO problem given by Eq.~(\ref{eq:general_qubo}). There are commercial QA devices available, like the ones offered by D-Wave Systems Inc.~\cite{dwave2019leap,dwave2020TechnicalDescription}, which have been shown to be able to solve reasonably-sized problems \cite{Pudenz2012QML, PerdomoOrtiz2019ReadinessQuantumOptimizationMethodsPUBOQUBO, Willsch2020QSVM, Willsch2021BenchmarkAdvantage}. Here we will focus exclusively on QA. 

A quantum annealer operates according to the adiabatic theorem \cite{born1928adiabatictheorem}. This theorem states that given a simple initial Hamiltonian with a known ground state and a problem Hamiltonian whose unknown ground state encodes the solution to the problem (e.g. our garden optimization QUBO problem Hamiltonian), by evolving the system according to the Schr\"odinger equation sufficiently slowly, we can ensure that the system stays in its ground state throughout the evolution from initial to problem Hamiltonian, leaving the system in the desired ground state which solves the problem \cite{KadowakiNishimori1998QuantumAnnealing,Fahri2000AdiabaticQuantumComputation}. On D-Wave quantum annealers, the parameter which controls the duration of this evolution process is the \emph{annealing time} (AT). This parameter defaults to $20\,\mu\mathrm s$ but can be set by the user to any value between $1\,\mu\mathrm s$ and $2000\,\mu\mathrm s$ \cite{dwave2020TechnicalDescription}.

A prerequisite for solving a problem on a QA device is that it is formulated in terms of the Ising model \cite{KadowakiNishimori1998QuantumAnnealing} or a QUBO problem such as Eq.~(\ref{eq:general_qubo}). Given that a formulation of the garden optimization problem in terms of a QUBO problem was provided in the previous section, we limit our discussion to this model. Beyond practical considerations on a per-problem basis like the possibility to perform simplifications when stating the problem using one model or the other, these two models are equivalent up to a trivial transformation of the domain of the binary variables from ${\{-1,+1\}}$ to ${\{0,1\}}$ or backwards.

Here we solve a set of garden optimization problems of increasing number of variables using a suite of QPUs as well as hybrid and classical software solvers offered by D-Wave Systems. In the following two sections, we review the samplers and solvers that we used for the benchmarks reported in this paper.  

\subsection{D-Wave QPUs}
\label{sec:qpus}

\begin{figure*}
    \centering
    \includegraphics[width=\textwidth]{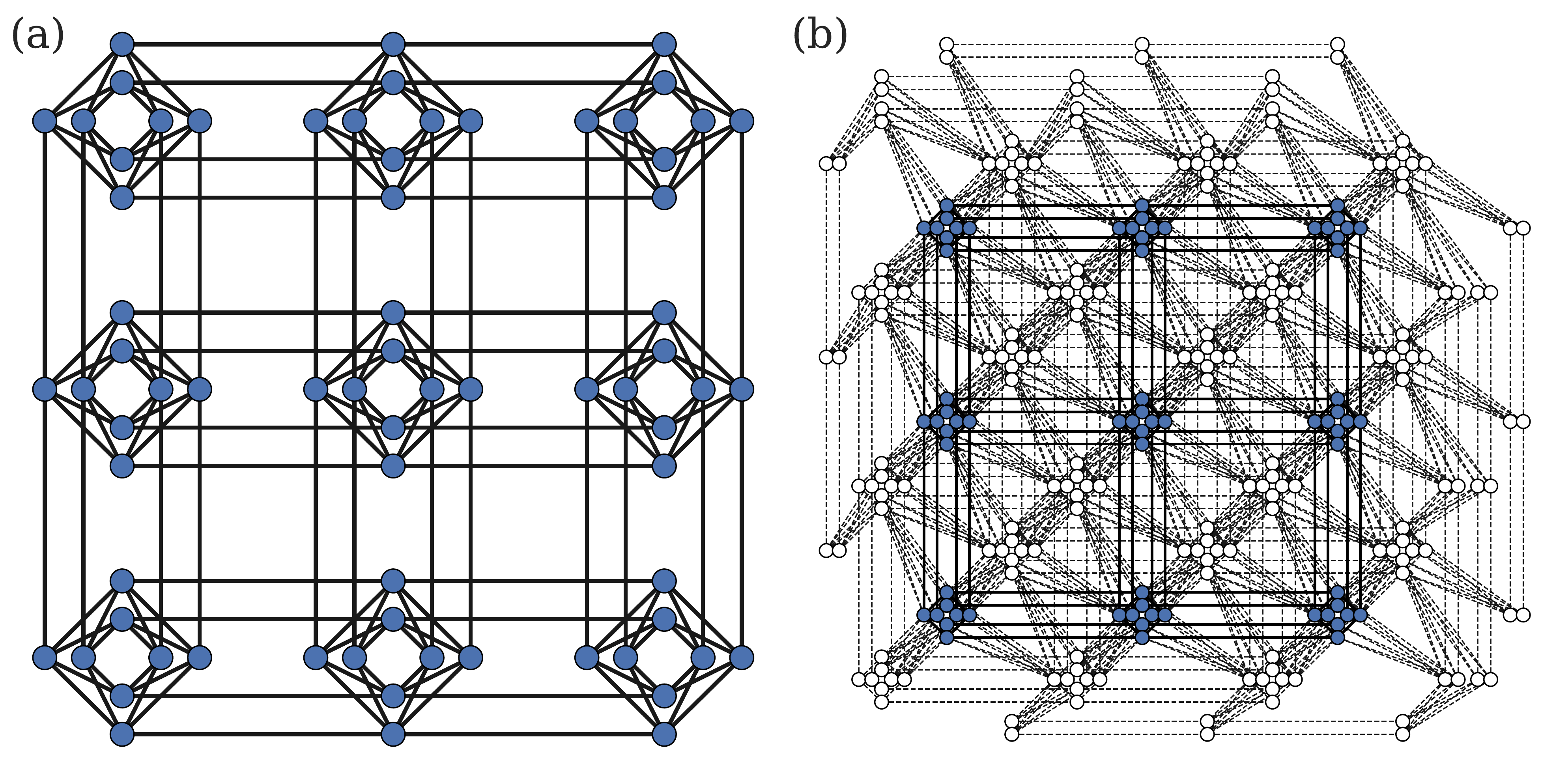}
    \caption{Graphs of Chimera and Pegasus topologies, where nodes and edges represent qubits and couplers, respectively. (a) Chimera graph of size 3 ($C_3$). (b) Pegasus graph of size 4 ($P_4$) where nodes are shown as white circles and edges as dashed lines, with a $C_3$ graph embedded in it, where nodes are shown as blue circles and edges as solid lines.}
    \label{fig:topologies}
\end{figure*}

D-Wave has offered access to a 2000+ qubit quantum annealer called \texttt{DW2000Q} since January 2017. Connecting the over 2000 qubits, this system features over 6000 couplers arranged in a Chimera graph of size 16, also known as $C_{16}$ \cite{Bunyk2014DWave}. In September 2020, a new system called \texttt{Advantage} was released, which features over 5000 qubits and over 35000 couplers arranged in a Pegasus topology of size 16 ($P_{16}$) \cite{dwave2020TechnicalDescription}. Besides the significant increase in the number of qubits, the most noteworthy improvement of the new generation was the qubit connectivity increasing from 6 to 15, which should allow the user to solve problems with more variables and denser connectivity \cite{dwave2020Advantage}. The characteristic form of the Chimera and the Pegasus graph is shown in Figs.~\ref{fig:topologies}(a) and (b), respectively.

A critical step in programming a quantum annealer is the embedding of the problem onto the \emph{working graph} of the chip \cite{Choi2008Embedding}. Typically, a small portion of the qubits and couplers in the chip are deactivated due to malfunction, so we define the working graph as the subset of the full-yield graph of the chip (described by the topology and size parameter, e.g.~$C_{16}$) which excludes the malfunctioning qubits and couplers. Finding an embedding onto the working graph of the chip (which allows us to run the problem on the quantum annealer) is generally a challenging problem in itself \cite{Choi2008Embedding, Okada2019EmbeddingLargerSubproblemsDWave}. It often relies on heuristic methods and requires several tries until a satisfactory embedding is found. 

In the special case that the graph of the QUBO problem can be directly mapped onto the working graph of the QPU, the embedding is a trivial step. However, most problems on a real-life scale do not meet this requirement and the embedding step becomes crucial. In order to overcome the limited connectivity of the QPU, the concepts of logical and physical qubits are introduced. When a direct embedding exists, the logical qubits (which represent the binary variables in the QUBO matrix) map directly onto the physical qubits of the QPU. However, when such an embedding cannot be found, logical qubits need to be represented on the QPU as ``chains'' of physical qubits. A qubit chain is a group of physical qubits that are coupled together strongly enough so that they behave as an effective, single logical qubit. This technique has the advantage of increasing the connectivity of the logical qubit since each of the participating physical qubits has its own couplers. The drawback is that this ``strong enough'' coupling strength has to be determined empirically on a per-embedding basis. If the coupling is too weak the chains break, which means that the logical qubits no longer behave as single units. In such a case, post-processing techniques (as for instance a majority vote) can be applied to determine a value for the logical qubit. However, a result obtained in this way can have a high energy in the original problem formulation. On the other hand, if the chain coupling strength is set too high, the couplers involved in keeping chains consistent will dominate over the couplers representing the original problem interactions. This effectively produces a new problem that is more concerned with keeping chains consistent, so that it no longer yields solutions to the original problem.

A reasonable starting point for optimizing the chain strength for a given embedded problem is the largest bias or coupling strength of the original QUBO problem. Therefore, for an easier comparison across problems, we define the \textit{relative chain strength} as 
\begin{align}
\label{eq:rcs}
	\mathrm{RCS} = \frac{\mathrm{ACS}}{\max\limits_{kk'} \lvert Q_{kk'}\rvert},
\end{align}
where $\mathrm{RCS}$ stands for relative chain strength, $\mathrm{ACS}$ is the (absolute) chain strength, and $Q_{kk'}$ is the QUBO matrix from Eq.~(\ref{eq:general_qubo}).

Moreover, a problem with denser connectivity requires longer chains, which in turn demands stronger chain coupling strengths since longer chains have a higher chance of breaking. Therefore, embeddings with fewer, shorter chains should be preferred. The embedding issue stresses the importance of the increased connectivity between the Chimera architecture on the \texttt{DW2000Q} and the Pegasus architecture on the new D-Wave \texttt{Advantage}. The Pegasus architecture should allow for equally or more compact embeddings than Chimera for any given problem. Also the increased number of qubits would allow us to run bigger problems that do not fit on the \texttt{DW2000Q} chip. 

\subsection{D-Wave hybrid solvers}
\label{sec:hybriddesc}

In addition to hardware QPUs like \texttt{DW2000Q} and \texttt{Advantage}, D-Wave Systems offers access to QUBO solvers following a hybrid approach. As we saw in the previous section, the current QPUs have a limited number of qubits which might not be enough to solve problems at a real-world scale. Hybrid solvers have been designed to overcome this limitation by classically partitioning the problem into sub-problems that are small enough to be solved on a QPU. The version 1 solver \texttt{HSSv1} was released in February 2020, using \texttt{DW2000Q} as the hardware backend to solve the sub-problems and being able to run problems of up to 10,000 variables. In September 2020, parallel to the release of the \texttt{Advantage} QPU, the version 2 solver \texttt{HSSv2} was released. The new hybrid solver relied on \texttt{Advantage} instead and was able to solve sparsely-connected problems of up to one million variables or fully-connected problems of up to 20,000 variables \cite{dwave2020AdvantageTechnologyUpdate}.

In order to perform a more comprehensive performance comparison of \texttt{HSSv1} and \texttt{HSSv2}, we have included in this study two purely classical QUBO solvers provided by the D-Wave Ocean SDK: \texttt{TabuSampler}, an MST2 multistart search algorithm \cite{Palubeckis2004MultistartTabuSearch}; and \texttt{QBSolv} \cite{qbsolv}, a partitioning algorithm which solves the sub-problems using \texttt{TabuSampler} as backend. We thus have two different hybrid classical/quantum partitioning solvers, a classical non-partitioning solver, and a classical partitioning solver. 

\blue{The hybrid solvers are proprietary software of D-Wave Systems Inc.~for which no information on their internal function is publicly available (see the official documentation on Leap Hybrid solvers \cite{DWaveSolversParameters}). As they do not offer tunable parameters (except for an optional timeout parameter), no parameter tuning was performed on any of the solvers. In other words, we consider all solvers as out-of-the-box solutions in their default mode of operation.}

\section{Results}
\label{sec:results}

In this section, we present the results obtained for the hardware samplers (\texttt{Advantage} and \texttt{DW2000Q}), as well as for the hybrid and software solvers (\texttt{HSSv1}, \texttt{HSSv2}, \texttt{TabuSampler} and \texttt{QBSolv}) in two separate subsections. For the hardware samplers we created four problem instances and performed a chain strength scan for several embeddings of each problem. Subsequently we chose the most successful embedding together with its optimal RCS value to perform an annealing time scan. These scans were used for comparing the performance of the hardware samplers. For the hybrid and software solvers we created 324 problem instances which we submitted to each solver. The Python 3 code used to generate these problem instances is available online as a Jupyter Notebook at Ref.~\onlinecite{GardenOnline}. Here we present the energy of the returned solutions as well as the execution times required to reach them. 

\subsection{Hardware samplers}
\label{qpus_results}

For the D-Wave QPUs \texttt{DW2000Q} and \texttt{Advantage} (see Sec.~\ref{sec:qpus}), we define a successful solution as one which fulfills all the constraints in Eqs.~(\ref{eq:const_pots})--(\ref{eq:const_size}) regardless of the quality of the placement of the plants in the garden (i.e., regardless of the contribution of the cost function Eq.~(\ref{eq:objectivefunction}) to the solution energy). \blue{We remark that there is no tolerance in the success of a solution, as we only count solutions as successful if they satisfy all constraints.} The \emph{success rate} is then defined as the ratio of successful solutions to the total number of samples produced. Defining the success rate in this way allows us to assess the quality of the solutions provided by a batch of samples in this scenario, where we do not know if we have found the ground state of the problem unless the energy of the sample is zero. An alternative metric which could be used to measure the success of a given embedded problem is the energy of the best sample, i.e., the \emph{lowest energy} (as considered in Sec.~\ref{hybrid_results} for comparing hybrid and classical solvers). This metric has the advantage over the success rate of taking into account not only the satisfaction of the constraints but also the quality of the placement of plants in the garden. \blue{However, for the hardware samplers evaluated here, the lowest energy turned out not to be a sensitive enough metric to tune the RCS and AT. The reason for this is  that among the valid solutions, the different samplers and embeddings often returned identical lowest energy results (data not shown)}.    

Garden problems of different size were created to compare the performance of the \texttt{DW2000Q} and \texttt{Advantage} systems at solving QUBO problems. Given the problem size limitations of the studied systems, the problem suite consists of four problems of 16, 36, 64, and 100 variables. We ensure that the created problems are satisfiable by setting $n$ to an even number (here $n \in \{4,6,8,10\}$) and sampling without replacement $n/2$ times from each of both the sets of big plants ($\{j : 0<=j<=t-1$ and $s_j=0\}$) and small plants ($\{j : 0<=j<=t-1$ and $s_j=1\}$) to obtain the list of plants to place in the garden. Sampling without replacement generates problems where there is at most one plant specimen of each species, so the number of variables in these problems is equal to $n^2$. 

For each of these problems, 10 embeddings were generated for each of the working graphs of \texttt{DW2000Q} (a subset of the $C_{16}$ graph) and \texttt{Advantage} (a subset of the $P_{16}$ graph) using the \blue{\texttt{find\_embedding} function of the} \texttt{minorminer} module included within the D-Wave Ocean SDK (see \cite{dwave2020TechnicalDescription} for more information on the working graphs \blue{and \cite{cai2014practical} for information on the embedding algorithm used}). These embeddings were successfully created for all problems except for the 100 variable problem on \texttt{DW2000Q}, since it was too big to fit on the \texttt{DW2000Q} chip. Therefore, for this problem, we only report results obtained on \texttt{Advantage}.

In order to study the influence of the additional couplers introduced by the enlarged connectivity of the Pegasus topology, the \texttt{DW2000Q} embeddings were mapped onto the Advantage graph (see Fig.~\ref{fig:topologies}(b) for a visualization of how the Chimera graph can be embedded onto the Pegasus graph, and by extension how a Chimera-embedded problem can be mapped onto a Pegasus graph). By this process, we obtain 10 new embeddings for each problem which could be used on the \texttt{DW2000Q} system in addition to the two original sets of 10 embeddings. In what follows, we refer to these Chimera embeddings used on the \texttt{Advantage} system as Advantage(Chimera) embeddings. Since the Advantage(Chimera) embeddings were not created specifically for \texttt{Advantage}, the compatibility of these embeddings with the working graph of the \texttt{Advantage} chip was checked prior to executing the problems, so that they could be replaced in the case that any of the involved qubits or couplers were not available.

\subsubsection{Chain strength scan}

\begin{figure}
    \centering
	\includegraphics[width=\linewidth]{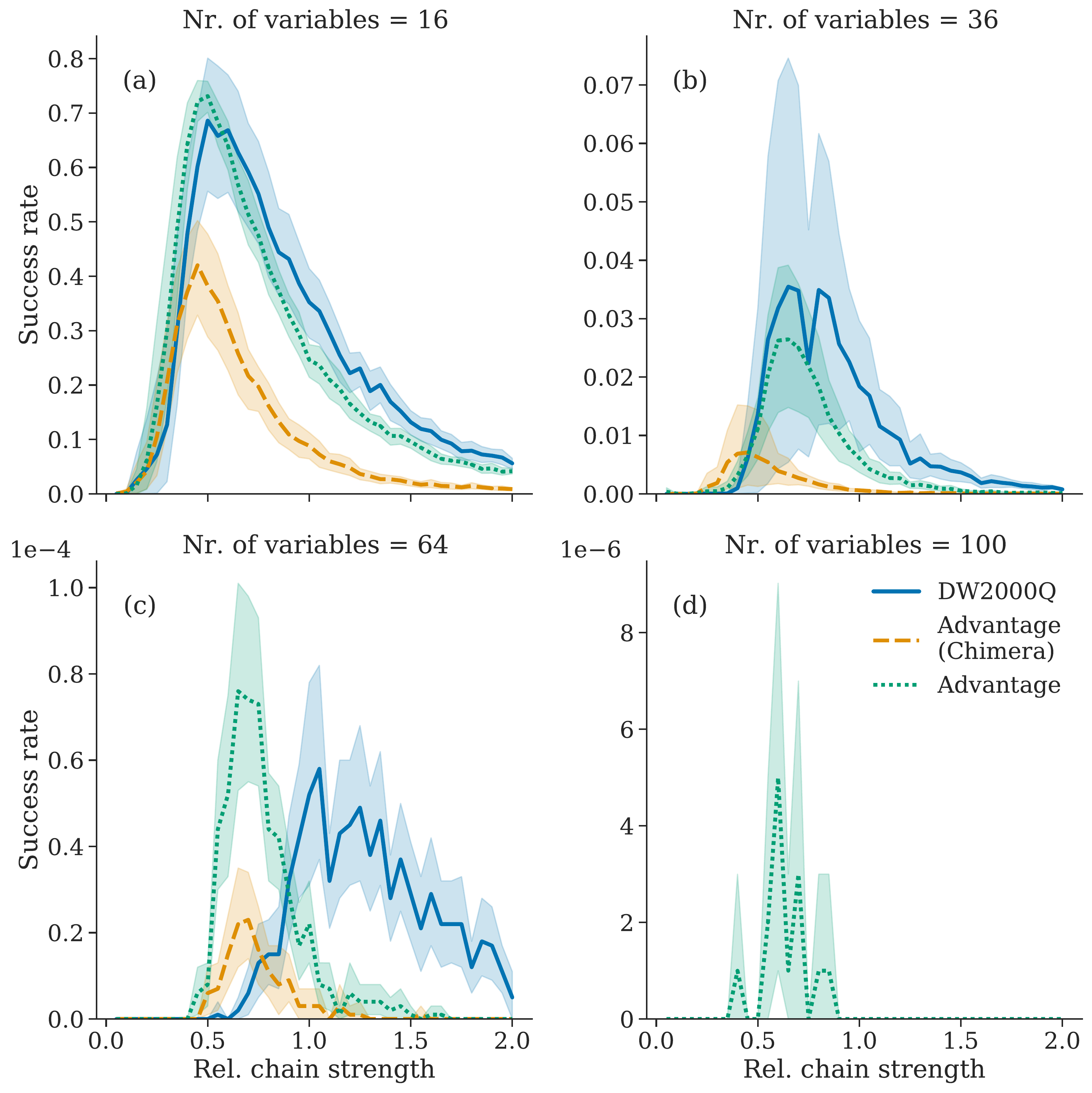}
    \caption{Success rate as a function of the $\mathrm{RCS}$ (see Eq.~(\ref{eq:rcs})) for four garden problems with increasing number of variables. For each problem size, 10 embeddings where generated for both $C_{16}$ and $P_{16}$ topologies and executed with $r$ samples. Full blue lines represent $C_{16}$-embedded problems executed on \texttt{DW2000Q}, dashed orange lines represent $C_{16}$-embedded problems executed on \texttt{Advantage} and dotted green lines represent $P_{16}$-embedded problems executed on \texttt{Advantage}, shown everywhere with 95\% confidence intervals. (a) 16-variable problem, with $r=10^4$ samples. (b) 36-variable problem, with $r=10^4$ samples. (c) 64-variable problem, with $r=10^5$ samples. (d) 100-variable problem, with $r=10^5$ samples. Note that in (d), only \texttt{Advantage} results are shown here since no embedding could be found for the $C_{16}$ graph.}
    \label{fig:rcs}
\end{figure}

The available sets of embeddings for the four problems were submitted to the respective systems 40 times with $\mathrm{RCS}$ values ranging from 0.05 to 2.00, increasing in steps of 0.05 (cf.~Eq.~(\ref{eq:rcs})). These sets of 40 jobs were used to scan the success rate as a function of the $\mathrm{RCS}$ for each embedding. The results of the $\mathrm{RCS}$ scan for all embeddings can be found in Fig.~\ref{fig:rcs}. All jobs in each scan produced $10^4$ samples for the $16$ and $36$ variable problems, and $10^5$ samples for the $64$ and $100$ variable problems. The number of samples in the latter problems was increased to produce sufficient statistics given the small success rates. 

The results in Fig.~\ref{fig:rcs} show that the success rate can be drastically improved by tuning the $\mathrm{RCS}$ for a given embedding. As the problem size increases, so do the chain lengths, leading to higher optimal $\mathrm{RCS}$ values required to stop the chains from breaking. This effect is more clearly visible on \texttt{DW2000Q}, possibly due to properties of this particular system, otherwise it would also be observable in the Advantage(Chimera) scans. For problems of up to 64 variables, for which Chimera embeddings were found, the performances of the \texttt{DW2000Q} and \texttt{Advantage} systems are comparable, whereas Advantage(Chimera) embeddings performed significantly worse. This difference in performance cannot be attributed to the embedding quality. Therefore, we conjecture that the additional unused couplers of the Pegasus architecture in the Advantage(Chimera) embedded problems when compared to the \texttt{DW2000Q} embedded problems play a role in lowering the success rate. For the 100 variable problem, only Pegasus embeddings could be found; and although the success rates are small, this indicates that the \texttt{Advantage} system is able to solve bigger problems due to its increased qubit number and connectivity. We also note that for the 36 variable problem, \texttt{DW2000Q} seems to outperform \texttt{Advantage}. This might be due to the fact that Chimera and Pegasus embeddings for small problems have similar chain lengths and the \texttt{DW2000Q} embeddings could have less unused couplers connected to qubits involved in the embeddings.

\subsubsection{Annealing time scan}

After performing the $\mathrm{RCS}$ scans presented in the previous section, we proceed to tuning another crucial parameter, the AT.  For every set of 10 embeddings used for the $\mathrm{RCS}$ scan in Fig.~(\ref{fig:rcs}), we pick the embedding which achieved the highest success rate for each problem and embedding type combination. For each picked embedding, we determine its optimal $\mathrm{RCS}$ value, and we proceed by performing an AT scan using the chosen embedding and $\mathrm{RCS}$ value combinations. It should be noted that coincidentally, the best \texttt{DW2000Q} embedding for each problem was the same as the best for Advantage(Chimera), although the optimal $\mathrm{RCS}$ was slightly lower in the Advantage(Chimera) embedding for problems of 36 and 64 variables and the same for the 16 variable problem.

Although both systems use a default AT of $20 \,{\mu}\mathrm s$, \texttt{DW2000Q} and \texttt{Advantage} allow the user to set the AT value between $1\,{\mu}\mathrm s$ and $2000\,{\mu}\mathrm s$. Therefore, the AT scan was performed using 20 values evenly spaced on a logarithmic scale in this range. All jobs in each scan produced $10^4$ samples for the $16$ and $36$ variable problems, $10^5$ samples for the $64$ variable problem and $10^6$ samples for the $100$ variable problem. The number of samples in the latter problems was increased to produce sufficient statistics given the small success rates. 

\begin{figure}
    \centering
	\includegraphics[width=\linewidth]{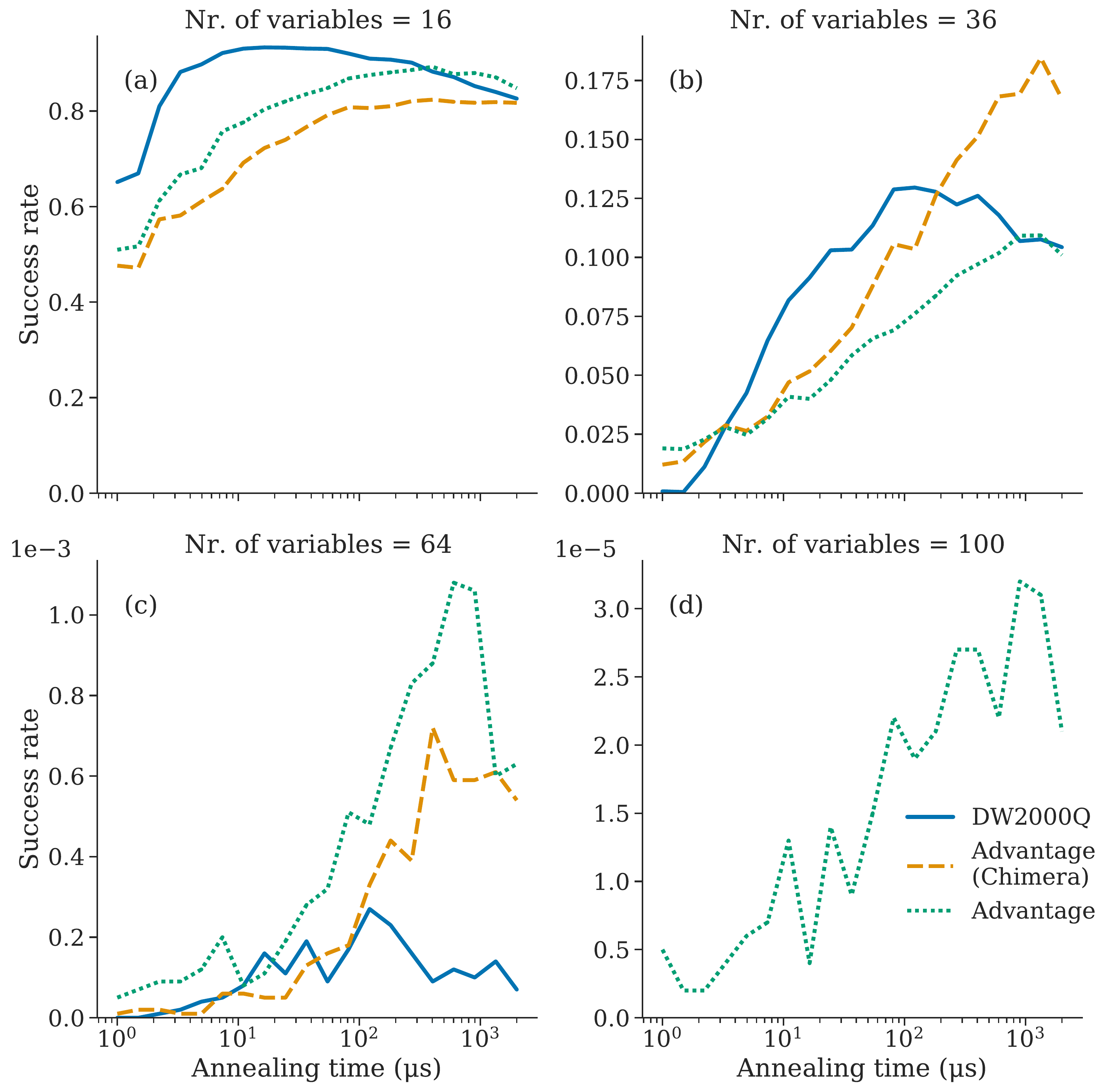}
    \caption{Success rate as a function of the annealing time for four garden problems of increasing number of variables. For each problem size, the most successful embedding from Fig.~\ref{fig:rcs} was selected along with its optimal chain strength and executed with $r$ samples. Full blue lines represent $C_{16}$-embedded problems executed on \texttt{DW2000Q}, dashed orange lines represent $C_{16}$-embedded problems executed on \texttt{Advantage} and dotted green lines represent $P_{16}$-embedded problems executed on \texttt{Advantage}. (a) 16-variable problem, with $r=10^4$ samples. (b) 36-variable problem, with $r=10^4$ samples. (c) 64-variable problem, with $r=10^5$ samples. (d) 100-variable problem, with $r=10^6$ samples. Note that in (d), only \texttt{Advantage} results are shown since no embedding could be found for the $C_{16}$ graph.}
    \label{fig:at}
\end{figure}

The results in Fig.~\ref{fig:at} show that the success rate can also be improved by tuning the annealing time for a given embedding. The general trend shows that higher AT values lead to higher success rates. However, since longer annealing times for a fixed number of samples result in longer execution times, it should be carefully considered for any given problem whether the increase in execution time resulting from setting a larger AT value would not be better spent in generating additional samples with shorter annealing times rather than fewer samples with longer annealing times. Comparing \texttt{Advantage} and Advantage(Chimera) against \texttt{DW2000Q}, we see that increasing the AT tends to be more consistently effective on the \texttt{Advantage} system. \blue{On \texttt{DW2000Q}, however, the success rate is found to decrease again for larger AT (see the blue curves in Fig.~\ref{fig:at}). This may be an effect caused by the environment (see \cite{Amin2015SearchingForQSQuasistaticAT} for more information).} It is possible that if the \texttt{Advantage} system accepted AT values beyond $2000 \,{\mu}\mathrm s$ we would observe similar behaviour, where the success rate decreases past some optimal AT value.   

\blue{
Figure~\ref{fig:at} also shows that the largest success rates in each case decrease with increasing problem size. One reason for this is that the qubit chain lengths increase as the number of variables grows. Another reason is that, although the energy gap is always between one or two by construction (see Eq.~\ref{eq:my_qubo}), larger problems may have a smaller effective energy gap due to the rescaling of the problem parameters into the range of available qubit and coupler strengths (see the auto-scaling procedure at \cite{DWaveSolversParameters}).} 

\subsection{Hybrid and software solvers}
\label{hybrid_results}

In this section, we compare the performance of the hybrid solvers \texttt{HSSv1} and \texttt{HSSv2} against the classical software solvers \texttt{TabuSampler} and \texttt{QBSolv} (see Sec.~\ref{sec:hybriddesc} for details on these solvers). The hybrid solvers were executed online on D-Wave servers and the classical solvers were executed on an Intel(R) Core(TM) i7-4770 CPU @ $3.40\,$GHz workstation with $32\,$GB of RAM. For this study, we generated increasingly bigger problems in an attempt to evaluate how the performance of the solvers behaves under increasing problem sizes. Each problem instance was created by fixing the number of pots $n$ to an even number and sampling with replacement the sets of available small and big plant species $n/2$ times each, so as to ensure that as many small plants as big ones were to be placed in the garden. Here we sample with replacement (as opposed to the problems created in Sec.~\ref{qpus_results}) to allow for more than one plant specimen per species. Thereby, we create problems with more pots than unique species defined in the companions matrix $C_{jj'}$ shown in Fig.~\ref{fig:companions}. The biggest problem which could be created given the available memory on the used workstation was for $n=684$ pots, with a total of $12312$ variables. Note that the biggest problems are beyond the problem size limit of \texttt{HSSv1} ($10000$ variables), but within the limit for \texttt{HSSv2} ($20000$ variables, fully connected). Each of the $342$ problems in this set was submitted to each of the studied solvers, so we can make a direct comparison of their performance. Here, rather than comparing the success rates as we did in the previous section, we compare the energy of the single sample returned by each of the solvers, as well as the execution times taken to return said sample.

\subsubsection{Energies}

\begin{figure}
    \centering
	\includegraphics[width=\linewidth]{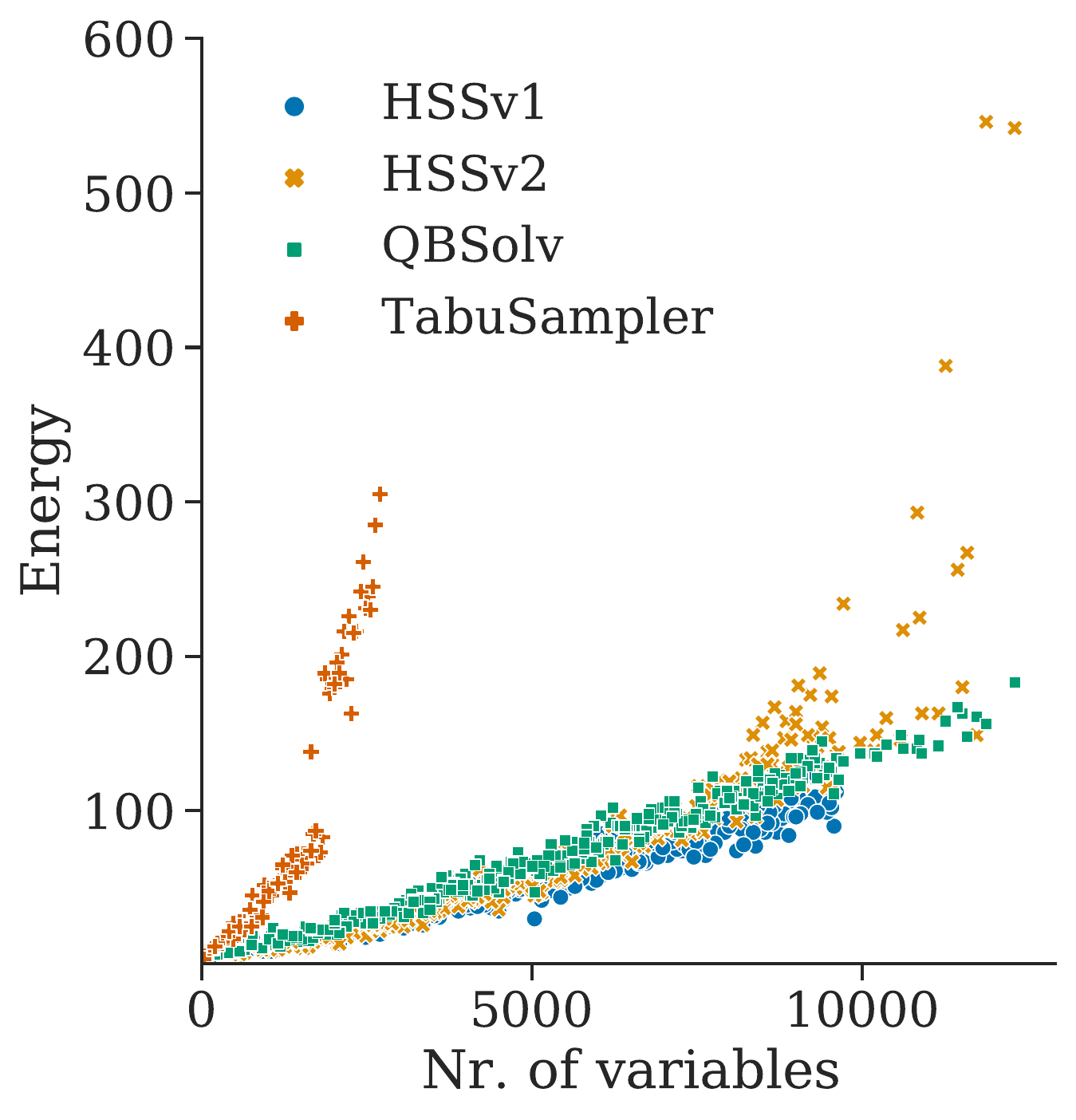}
    \caption{Lowest energy returned by \texttt{HSSv1} (blue circle markers), \texttt{HSSv2} (yellow X markers), \texttt{QBSolv} (green square markers) and \texttt{TabuSampler} (red cross markers) for garden problems of up to 12312 variables. Note that \texttt{TabuSampler} energy results for problems beyond 2500 variables are not shown since they were several orders of magnitude larger than the rest.}
    \label{fig:energies_full}
\end{figure}

Figure \ref{fig:energies_full} shows the energies returned by each of the solvers for the generated set of garden problems. \texttt{TabuSampler} seems to perform significantly worse than the others. This is probably due to the fact that it is the only solver that does not partition the problem into separately solved sub-problems, but attempts to solve the complete problem at once. The energy results for this solver were at least three orders of magnitude larger than the rest for problems beyond $2500$ variables. \texttt{QBSolv} displays a rather predictable linear trend for increasing problem size in the studied range of problems. For problems below $5000$ variables, \texttt{HSSv2}, \texttt{HSSv1} and \texttt{QBSolv} show very similar performance, with \texttt{QBSolv} returning slightly worse results. Above $5000$ variables, the differences become clearer, with \texttt{HSSv1} outperforming all other solvers up until the $10000$ variable barrier, after which no results are available for this solver. Surprisingly, as the problem size increases, \texttt{HSSv2} seems to perform comparatively worse than \texttt{QBSolv} and especially \texttt{HSSv1}. Additionally, the spread of the energies returned by \texttt{HSSv2} increases compared to that of the other solvers. 

\begin{figure}
    \centering
	\includegraphics[width=\linewidth]{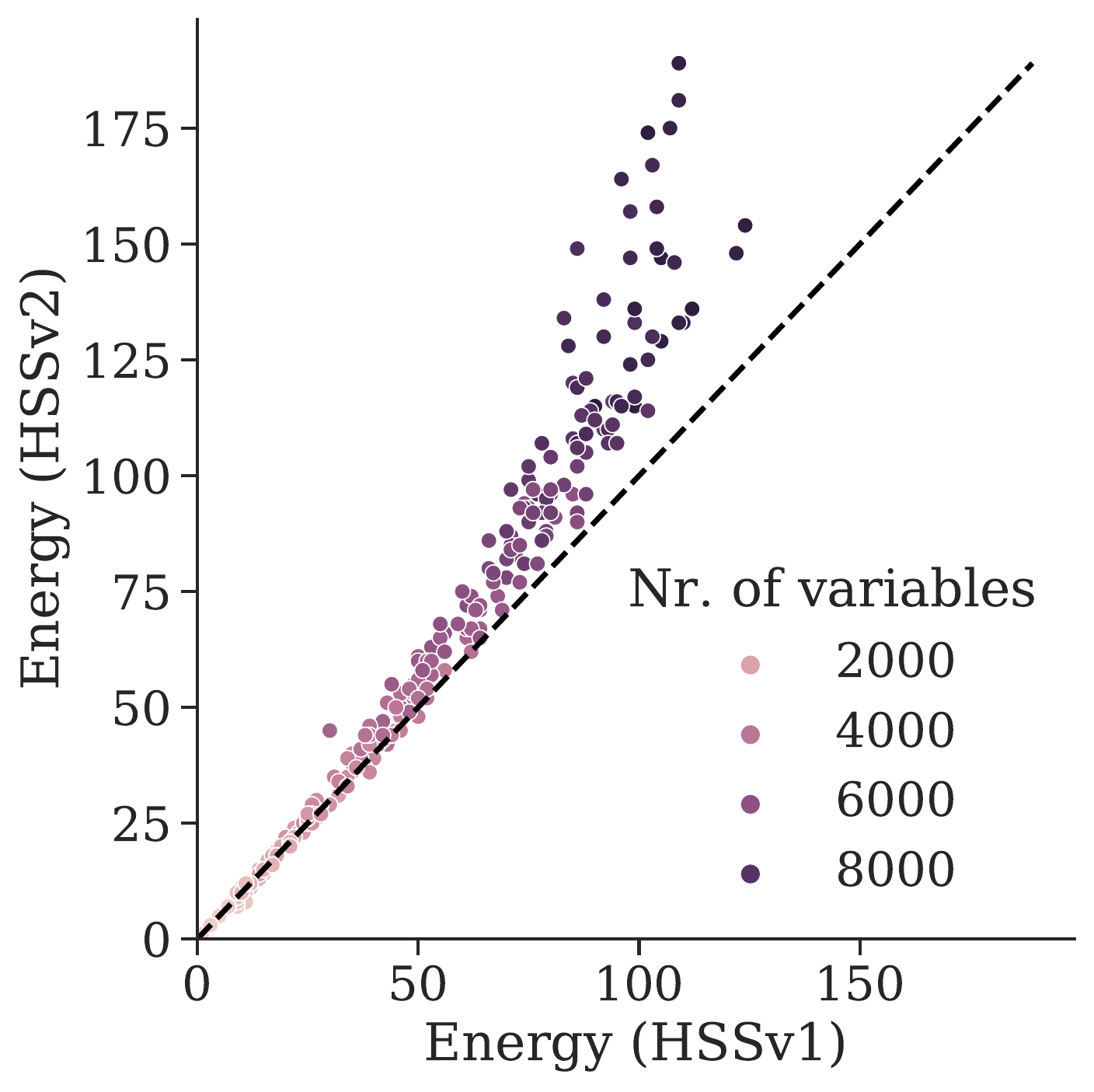}
    \caption{Direct comparison of the energies returned by \texttt{HSSv1} (x axis) and \texttt{HSSv2} (y axis) for the set of problems from Fig.~\ref{fig:energies_full}. The brightness of the colored points is used  to indicate the number of variables of each problem, with darker points representing larger problems. The diagonal dashed line is a guide to the eye to show the separation between the region where \texttt{HSSv1} outperforms \texttt{HSSv2} (upper left half) and vice versa (lower right half). Note that problems with more than 10000 variables are omitted since they could not be submitted to \texttt{HSSv1}.}
    \label{fig:energies_direct}
\end{figure}

To study this trend in more detail, a direct comparison of the energies returned by \texttt{HSSv1} and \texttt{HSSv2} for each problem is shown in Fig.~\ref{fig:energies_direct}. Here we plot the energy of both \texttt{HSSv1} and \texttt{HSSv2} on the x and y axes, respectively, and use the colour dimension to represent the number of variables of each problem represented by a point. For problems of up to $2000$ variables there is no visible difference between the solvers. However, as problems increase beyond this number of variables, the gap in the energies returned by each solver increases in favour of \texttt{HSSv1}. It should be noted that the results shown here should be interpreted exclusively within the scope of the studied garden optimization problems, and further evaluation involving different problem classes would be required to reach a general conclusion regarding the performance of these solvers.

\subsubsection{Execution times}

\begin{figure}
    \centering
    \includegraphics[width=\linewidth]{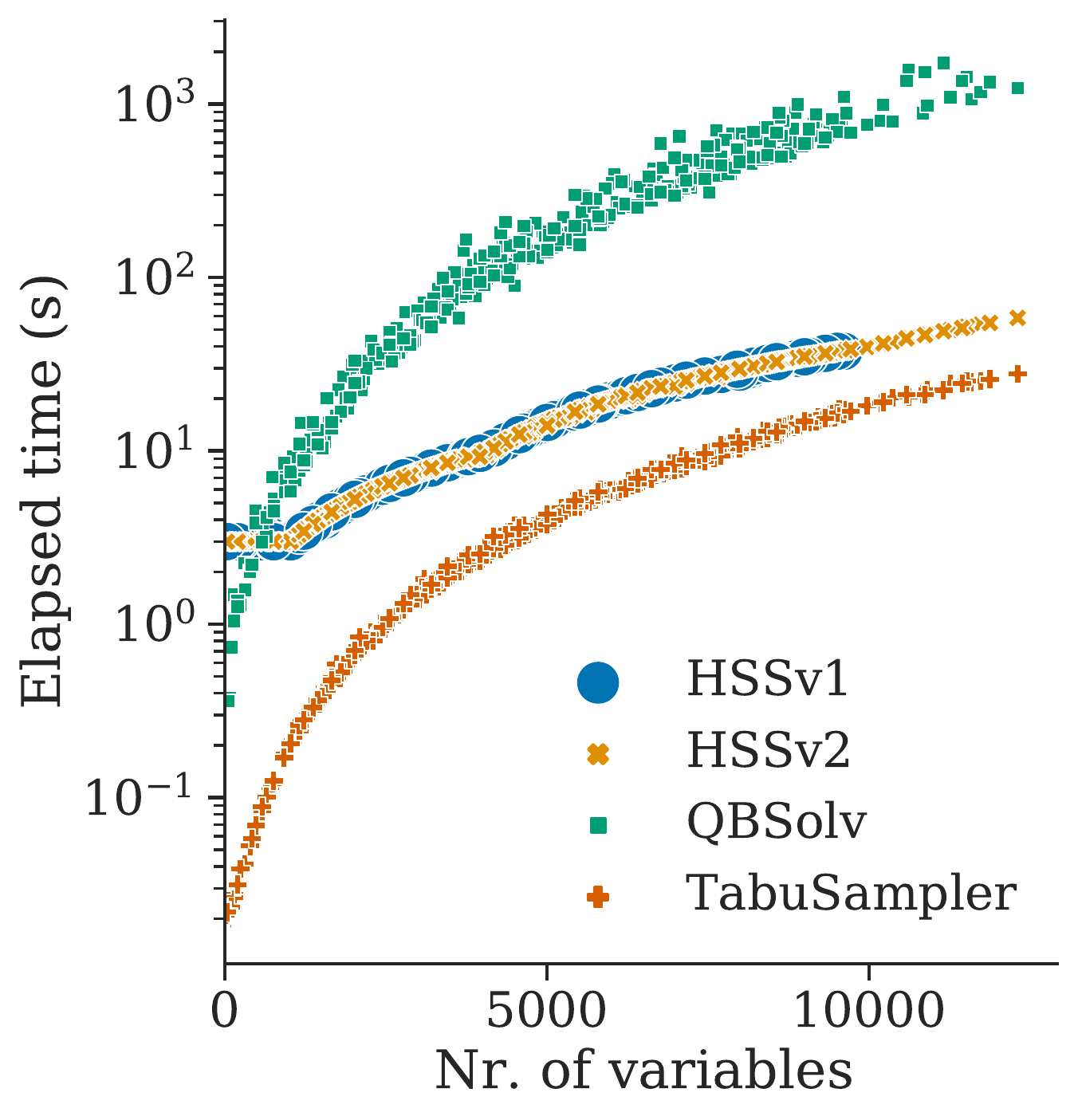}
    \caption{Execution times of \texttt{HSSv1} (blue circle markers), \texttt{HSSv2} (yellow X markers), \texttt{QBSolv} (green square markers) and \texttt{TabuSampler} (red cross markers) for garden problems of up to 12312 variables. Note that the different size of the markers does not have any special meaning; simply markers for \texttt{HSSv1} have been enlarged for visibility due to their execution times being identical to those of \texttt{HSSv2}.}
    \label{fig:runtimes}
\end{figure}

A comparison of the solvers would be incomplete if the execution times were not taken into consideration. Figure \ref{fig:runtimes} shows the time in seconds employed by every solver to solve each of the problems in the generated set. \texttt{TabuSampler} is the fastest among all tested solvers, although as it was shown in Fig.~\ref{fig:energies_full}, its returned energies were much worse than the others. \texttt{HSSv1} and \texttt{HSSv2} have identical execution times up to $10000$ variables. Note that the minimum execution time for \texttt{HSSv1} and \texttt{HSSv2} is $3\,\mathrm s$ independent of the problem size. Hence the execution time is constant for these two solvers for the smaller problems. Among all solvers, \texttt{QBSolv} is the slowest, with execution times at least one order magnitude larger than the hybrid solvers. Considering this together with the energy results shown in Fig.~\ref{fig:energies_full}, one might conclude that \texttt{HSSv1} is the most competitive among the tested solvers if the problem size does not exceed its limit of $10000$ variables. For problems above this threshold, however, it is unclear whether the shorter running times and higher energies of \texttt{HSSv2} should be preferred over the longer running times but closer to optimal solutions of \texttt{QBSolv}.

\section{Conclusion}
\label{sec:conclusion}

\blue{
In this paper we have presented garden optimization problems that are present in companion planting, as a class of QUBO problems suitable for benchmarking quantum annealing systems over a wide range of problem and system sizes. The scalability of the garden optimization problems has proven flexible enough to benchmark hardware samplers as well as hybrid and classical solvers, whose maximum problem size limits are around two orders of magnitude larger than for the currently available hardware samplers. We have reported results of benchmarking D-Wave hardware samplers \texttt{Advantage} and \texttt{DW2000Q}, as well as hybrid and classical solvers \texttt{HSSv1}, \texttt{HSSv2}, \texttt{TabuSampler} and \texttt{QBSolv}. To the authors' knowledge these have been the first benchmarking results of \texttt{Advantage} and \texttt{HSSv2} developed by researchers independent of D-Wave Systems. Nevertheless, we remark that before generalizing the results shown here, it would be good to verify the general trends by benchmarking these systems using other classes of problems. For similar studies see~\cite{Willsch2021BenchmarkAdvantage, Willsch2021JUQCSGQAOA, cohen2020picking, kuramata2020larger, birdal2021QuantumSync, Bhatia2021PerformanceAnalysisQSVM, fox2021bio, rahman2021su2}.

Regarding the results for the hardware samplers (see Sec.~\ref{qpus_results}) we can conclude that the recently released \texttt{Advantage} system is capable of embedding and solving garden optimization problems of $100$ variables, which is far beyond the reach of the previous generation of D-Wave systems, \texttt{DW2000Q}. This is due to the increase in the number of qubits and couplers in the newer system. For the evaluated problems with less than $100$ variables we did not observe significant differences in the performance of both systems. The results for Advantage(Chimera), i.e., a Chimera embedding used on \texttt{Advantage} (cf.~Fig.~\ref{fig:topologies}), suggest that the additional couplers of \texttt{Advantage}'s Pegasus architecture with respect to \texttt{DW2000Q}'s Chimera architecture can negatively impact the success rate if these couplers are not used in the embedding. However, these additional couplers allow for equally and often more compact embeddings for a fixed problem, so they should be beneficial when solving embedded problems, especially when the problems have more variables or denser connectivity.

The tunings of the relative chain strength (RCS) and annealing time (AT) values shown in Sec.~\ref{qpus_results} were performed sequentially, by first finding the optimal RCS value for the best embedding in each system and, after fixing these, tuning the AT value. Ideally, the optimal RCS and AT values should be searched simultaneously since there is no guarantee that these values are independent from each other. However, the computational cost of scanning a two-dimensional space for each of the available embeddings compared to scanning the parameters sequentially is too large to properly address this limitation here. For this reason, and also since the RCS scan values had a bigger impact on the success rate than the AT scan, we chose to perform the RCS scan first.

Regarding the results for the hybrid and classical solvers (see Sec.~\ref{hybrid_results}) we note that \texttt{HSSv1} is the most competitive among the tested solvers as long as the problem does not exceed $10000$ variables, since it has the shortest execution times together with \texttt{HSSv2} but returns lower energy results than the other solvers. Problems beyond $10000$ variables cannot be submitted to \texttt{HSSv1}. For the largest problems it is unclear whether \texttt{QBSolv} or \texttt{HSSv2} performs best: \texttt{QBSolv} has longer execution times but returns results with lower energies, whereas \texttt{HSSv2} is faster but the results returned have higher energies. 

Whereas the problem size limits and the performance behaviour approaching these limits could be tested for \texttt{HSSv1} with the generated set of problems, it would have been interesting to also include such results for \texttt{HSSv2}, \texttt{TabuSampler} and \texttt{QBSolv}. We leave the exploration of even larger problem instances to approach the size limits of \texttt{HSSv2} for future work.
}

\begin{acknowledgments}
We thank Madita Willsch, Fengping Jin, Manpreet Jattana and Hans De Raedt for helpful discussions and for proofreading the manuscript. We especially thank Madita Willsch for spotting an elusive mistake in the QUBO formulation. We thank D-Wave Systems for early access to an Advantage\textsuperscript{TM} system and the new hybrid solver service in Leap during the Advantage beta program. We gratefully acknowledge the J\"ulich Supercomputing Centre  for funding this project by providing additional computing time through the J\"ulich UNified Infrastructure for Quantum computing (JUNIQ) on the D-Wave quantum annealer.
D.W. acknowledges support from the project JUNIQ that has received funding from the German Federal Ministry of Education and Research (BMBF) and the Ministry of Culture and Science of the State of North Rhine-Westphalia. C.G. acknowledges support from the project OpenSuperQ (820363) of the European Quantum Flagship.
\end{acknowledgments}

\appendix

\section{Explicit QUBO expression}
\label{app:expanded_qubo}

In this appendix, we give the final QUBO expression of the garden optimization problem. It can be obtained from Eq.~(\ref{eq:my_qubo}) by multiplying out the squares and replacing the linear terms by quadratic terms (using the fact that for Boolean variables, $x_{ij} = x_{ij}^2$). Performing these steps, we find
\begin{align}
    	\label{eq:expanded_qubo}
        & \min_{x_{ij}\in\{0,1\}}\Bigg\{\\
        & t^2 \sum\limits_{i,i'=0}^{n-1}J_{ii'} 
        + \sum\limits_{i=0}^{n-1}
        \sum\limits_{i'=0}^{n-1}
        \sum\limits_{j=0}^{t-1}
        \sum\limits_{j'=0}^{t-1}
        x_{ij} J_{ii'} C_{jj'} x_{i'j'}\\
        +&\lambda_1 n
        + \lambda_1 \sum\limits_{i=0}^{n-1}
        \left( 
        -\sum\limits_{j=0}^{t-1} x_{ij}^2 
        + \sum\limits_{j=0}^{t-1} 
        \sum\limits_{j'<j}^{t-1} 2 x_{ij} x_{ij'}
        \right) \\
        +&\lambda_2 \sum\limits_{j=0}^{t-1} c_{j}^2
        + \lambda_2 \sum\limits_{j=0}^{t-1}
        \left[
        \left(1-2c_j\right) 
        \sum\limits_{i=0}^{n-1} x_{ij}^2
        + \sum\limits_{i=0}^{n-1}
        \sum\limits_{i'<i}^{n-1}
        2 x_{ij} x_{i'j}
        \right]\\
        +&\lambda_3
        \sum\limits_{i=0}^{n-1}\sum\limits_{j=0}^{t-1} \left(i\%2 - s_j\right)^2 x_{ij}^2
        \Bigg\}.
\end{align}
Note that the expression contains only constant and quadratic terms, and each quadratic term is a contribution of the form $c_{iji'j'} x_{ij} x_{i'j'}$. Thus, we obtain the values $Q_{kk'}$ (or $Q_{k'k}$ if $k'<k$) of the QUBO matrix in Eq.~(\ref{eq:general_qubo}) for $k=ti+j$ and $k'=ti'+j'$
by going through the terms and summing up the respective coefficients $c_{iji'j'}$. See the \texttt{build\_bqm} function in the example Jupyter Notebook available at Ref.~\onlinecite{GardenOnline} for a demonstration of how these coefficients are summed up explicitly.

\bibliographystyle{apsrev4-2custom}
\bibliography{references}

\end{document}